\documentclass[sigconf]{acmart}
\AtBeginDocument{%
  \providecommand\BibTeX{{%
    \normalfont B\kern-0.5em{\scshape i\kern-0.25em b}\kern-0.8em\TeX}}}

\copyrightyear{2021}
\acmYear{2021}
\setcopyright{acmlicensed}\acmConference[MM '21]{Proceedings of the 29th ACM International Conference on Multimedia}{October 20--24, 2021}{Virtual Event, China}
\acmBooktitle{Proceedings of the 29th ACM International Conference on Multimedia (MM '21), October 20--24, 2021, Virtual Event, China}
\acmPrice{15.00}
\acmDOI{10.1145/3474085.3475700}
\acmISBN{978-1-4503-8651-7/21/10}

\usepackage{hyperref}
\usepackage{ulem}
\usepackage{subfig}
\usepackage{balance}

\begin{document}
\fancyhead{}
\title{Armor: A Benchmark for Meta-evaluation of Artificial Music}


\author{Songhe Wang}
\authornote{Both authors contributed equally to this research.}
\email{songhe17@live.unc.edu}
\affiliation{%
  \institution{University of North Carolina at Chapel Hill}
  \streetaddress{}
  \city{Chapel Hill}
  \state{North Carolina}
  \country{USA}
  \postcode{27514}
}

\author{Zheng Bao}
\authornotemark[1]
\email{zhengbao@live.unc.edu}
\affiliation{
  \institution{University of North Carolina at Chapel Hill}
  \streetaddress{}
  \city{Chapel Hill}
  \state{North Carolina}
  \country{USA}
  \postcode{27514}
}

\author{Jingtong E}
\email{jingtong@live.unc.edu}
\affiliation{%
  \institution{University of North Carolina at Chapel Hill}
  \streetaddress{}
  \city{Chapel Hill}
  \state{North Carolina}
  \country{USA}
  \postcode{27514}
}


\begin{abstract}
Objective evaluation (OE) is essential to artificial music, but it's often very hard to determine the quality of OEs. Hitherto, subjective evaluation (SE) remains reliable and prevailing but suffers inevitable disadvantages that OEs may overcome. Therefore, a meta-evaluation system is necessary for designers to test the effectiveness of OEs. In this paper, we present Armor, a complex and cross-domain benchmark dataset that serves for this purpose. Since OEs should correlate with human judgment, we provide music as test cases for OEs and human judgment scores as touchstones. We also provide two meta-evaluation scenarios and their corresponding testing methods to assess the effectiveness of OEs. To the best of our knowledge, Armor is the first comprehensive and rigorous framework that future works could follow, take example by, and improve upon for the task of evaluating computer-generated music and the field of computational music as a whole. By analyzing different OE methods on our dataset, we observe that there is still a huge gap between SE and OE, meaning that hard-coded algorithms are far from catching human's judgment to the music. 
\end{abstract}


\ccsdesc[500]{Applied computing~Sound and music computing}

\keywords{datasets, neural model, music evaluation, music information retrieval}


\maketitle

\begin{figure}%
    \centering
    \subfloat[\centering Continuation collection and comparison task]{{\includegraphics[width=9cm]{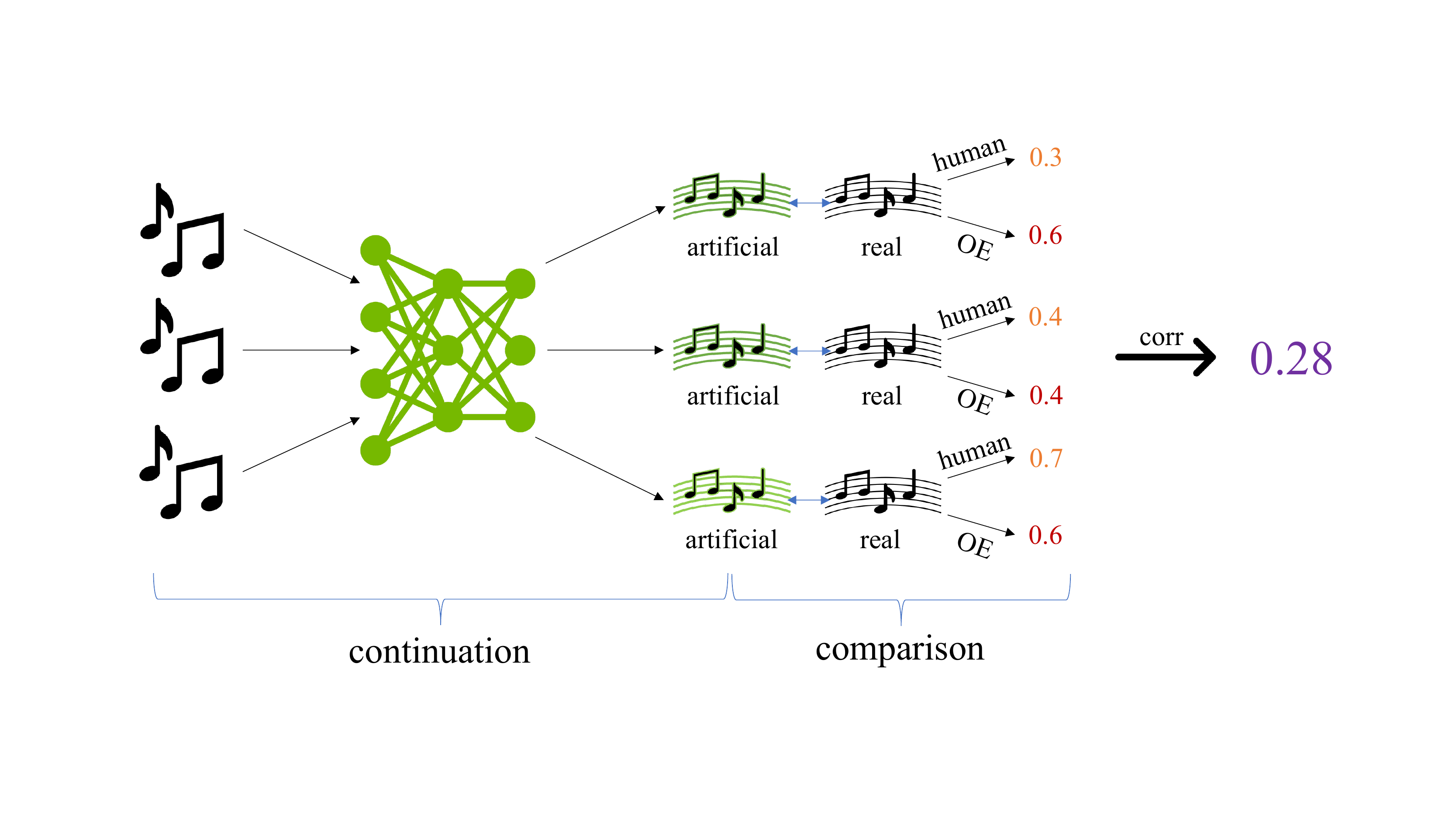} }}%
    \qquad
    \subfloat[\centering Music generated from scratch and distinguishing task]{{\includegraphics[width=9cm]{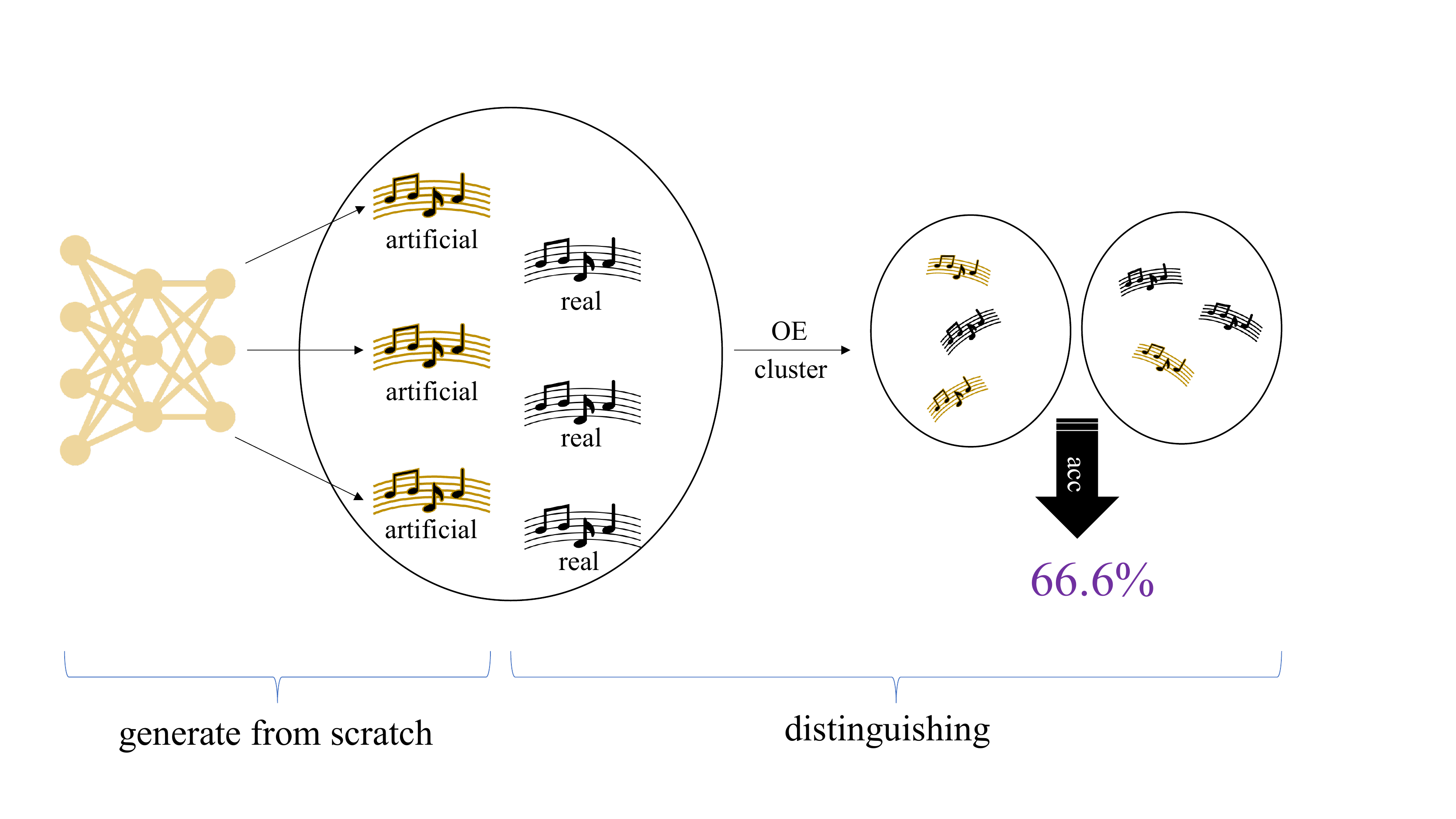} }}%
    \caption{An overview of Armor}%
    \label{fig:example}%
\end{figure}

\section{Introduction} 

Normally, there are two ways of evaluating the generated music: \textit{subjective} evaluation (SE) and \textit{objective} evaluation (OE). Traditionally, music was composed solely by well-trained musicians and evaluated by people with distinct understandings of music. Some define musical quality in compositional complexity, while some favor refreshing forms of arrangements with originality. Essentially, this demonstrates an SE, which requires respondents to listen to the generated music and evaluate the quality of the music based on their perception and music taste. Certainly, proposing an evaluation algorithm to convey human's tastes to music was never conceived contemporarily. 

\begin{table*}[t]
\centering
\begin{tabular}{lccc}
\toprule
Model & Objective evaluation & Subjective evaluation\\
\midrule
REMI \cite{REMI} & Newly proposed & Yes\\
SING \cite{defossez2018sing} & Solubility, Synthetizability, Druglikeliness & Yes\\
MIDI-Sandwich2 \cite{liang2019midisandwich2} & Not presented & Yes\\
LakhNES \cite{donahue2019lakhnes} & Perplexity & Yes\\
DeepBach \cite{hadjeres2017deepbach} & Not presented & Yes\\

\bottomrule
\end{tabular}
\caption{Ununiform usage of evaluation metrics for music generation models.}
\label{tab:evaluationmetricssituation}

\end{table*}

However, with the recent development of music generation boosted by deep learning \cite{hadjeres2017deepbach,liang2019midisandwich2, huang2020pop,donahue2019lakhnes,dong2017musegan, dong2018convolutional,guimaraes2018objectivereinforced,yang2017midinet}, the emergence of artificial music grants necessity for OE algorithm to evaluate artificial music works because when it comes to SE, it has several major issues: 1. it is extensively labor-intensive. 2. it is difficult to design labeling procedures that could avoid bias from different music tastes and perceptions of different people. 3. it is unlikely to obtain accurate feedback when a model is training or tuning. Deep learning models often require tuning or even modifying the model structures; therefore, it is crucial to acquire feedback on the quality of the music during the process of tuning and modifying. However, it is unlikely for the researchers to conduct SEs simultaneously while training a model. Therefore, OE methods are often used in music generation research papers to reflect the quality of generated music since it does not require human labor and could provide a relatively consistent standard to all the music. However, a well-accepted and universal OE metric has yet to emerge, and different models utilize different OE metrics (see Table \hyperref[tab:evaluationmetricssituation]{1}), which produces great difficulties for model comparison and, consequently, the guidance for future model construction. A meta-evaluation benchmark to test these metrics' effectiveness is in urgent demand to pave the way for better OE metrics.

A metric that gives more value to human judgments is the agreement among evaluators. \cite{Amabile1982SocialPO} In psychology, Cronbach's alpha \cite{Cronbach}, for instance, is the most common measure of internal consistency or reliability. Similarly, when it comes to evaluating artificial music, we consider SE the yardstick: OE should aim to yield results as similar as SE does since subjective method could accurately manifest human's judgments to music. According to this principle, we build \textit{Armor}\footnote{In terms of nomenclature, we would like the name of our dataset to symbolize our research’s topic of \textit{Auto MIR} (Automatic Music Information Retrieval) while taking the pronunciation of A-MIR as Armor. We hope that this dataset and the work as a whole would defend academic rigor, just like armor.}, a benchmark dataset that consists of two evaluation tasks: comparison and distinguishing. For both tasks, we provide music test cases and corresponding human-labeled scores to push OEs to minimize the gap between their results and human judgment. For the comparison task, an evaluation method should compare the generated music to the original one, tell how similar the two music pieces are, and give each pair a score. To obtain the generated music, we feed the generative model with an excerpt of original music, the prompt, and request it to write the continuation of the prompt or to complete the music. Through the above processes, the generated music would share some similarities with the original one. Another task is distinguishing, which requires the evaluation metric to determine whether music is human-composed or artificial and give a score on a corpus level. We first feed the generative model with random noise to generate a piece of music from scratch. We then manually classify the generated music and add the same number of human-composed music excerpts into the corpus according to the genre distribution of generated music. As a whole, Figure \hyperref[fig:example]{1} provides an overview of Armor. We choose the five most representative music generation models of recent years to perform the above generation tasks. We will discuss the specific reasons we choose these models in Section \hyperref[sec:models]{3.1}. 

A good evaluation metric should be \textit{universal} and \textit{continuous}. The universal evaluation suggests that the evaluation metric should give a fair evaluation score at different domains, and here we can refer to different music genres. Continuous evaluation suggests that the evaluation metric should not only assess complex music, but also judge music with simple structures and vice versa. Therefore, we design \textit{Armor} according to these two principles. First, we incorporate music with $21$ different genres into our corpus to diversify our dataset. Second, we deliberately choose models that could generate music with simple to complex structures. To show the new challenges our collected dataset brought to current OE metrics, we applied several commonly used OE metrics to our dataset. We found that none of the scores these metrics provide correlates well with the human-labeled scores.

Our contributions are: 1) We introduce a novel cross-domain dataset to test the effectiveness of OE metrics at two different scenarios. To our knowledge, \textit{Armor} is the first systematically collected benchmark dataset that could test the effectiveness of music evaluation metrics  2) We perform detailed dataset analysis and shed light on how different models, genres and number of tracks would affect human evaluators' performance. 3) We applied current OE metrics to our dataset and found out that they are unable to match human's judgment to music and thus propose our conjecture on how to build a good music OE metric based on our insight gained from the experiments.

\section{Related work}
Machine generated music has been constantly evolving in the past decades, along with numerous evaluation metrics to judge the quality of the generated music. There are numerous metrics based on different approaches, including but not limited to prediction and accuracy using statistics models, feature extraction, as well as listening tests. In general, evaluation metrics could be divided into two main categories: SE metrics and OE metrics. 

\subsection{SE metrics}
SE metrics include a series of listening tests that involves human judgments. Turing Test \cite{turing}, which asks respondents to decide whether the music they listen to is machine-generated or human-composed, is successfully implemented as an evaluation metric in audio synthesis \cite{hadjeres2017deepbach,cifka2019supervised,miranda2000music,Agarwala2017MusicCU,Liang2017AutomaticSC}. The Mean Opinion Score is another method of creating a collective evaluation based on human judgments. It requires participants to rate a piece of music on a scale from 1 to 5; the higher the score indicates, the better the music \cite{defossez2018sing,haque2018conditional}. In addition, side-by-side evaluation, which asks participants to rate the generated piece over the ground truth on a scale from -1 to 1, depending on whether the generated piece is better than the original piece, is also commonly used in evaluating generated music \cite{haque2018conditional}. However, these listening tests often have the risk of overestimating the understanding or accuracy of human judgments \cite{Ariza}.
\subsection{Non-feature based OE metrics}
OE metrics on the other hand, require no human-judgments. BLEU (Bilingual Evaluation Understudy Score) \cite{papineni-etal-2002-bleu} computes the similarity of consecutive segments between sequences and was first applied on the evaluation of machine translation. Later it was applied to music generation with graph neural network \cite{Jeong_Kwon_Kim_Nam_2019}. Another significant aspect to focus on is the area between two monotone chains. The minimized area between two chains could determine their similarity as well \cite{article}. In \cite{Jeong_Kwon_Kim_Nam_2019,barry2018style}, the Kullback-Leibler(KL) divergence of the Inter-Onset interval length was calculated to measure the similarity of the rhythm expressed in the source audio and the generated piece. 
\subsection{Feature based OE metrics}
There are other types of OE metrics that are based on extracting features of a piece of music as well. Variable-Markov Oracle \cite{Wang_Dubnov_2021} detects repeated patterns in a given subset of generated music, therefore determines the similarity between the two pieces. In addition, by extracting feature vectors that could identify a piece of music such as centroid, zero crossing rate, and key clarity, the similarity between two pieces of music could be determined using a supervised model \cite{inproceedings}.  Based on useful information from the texture of music, mel spectrogram is another way of evaluating generated music \cite{barry2018style}. There are plenty of music generation models emerging as the evaluation metrics are becoming more effective.

\section{Dataset collection}

 In Armor, all our human-composed music pieces are from Lakh \cite{raffel2016lakh}, a large-scale dataset that consists 176,581 unique MIDI files with \textit{21} genres, which provides a strong support to our dataset collection. In order to encourage universal and continuous OE metrics, a benchmark dataset like Armor should have music with diverse genres, single track pianoroll and multi-track polyphonic music and music with simple structures and complex structures. We carefully design \textit{Armor} to fulfill these requirements: Armor consists of MIDI files from each of the 21 genres presented in Lakh but follows a different distribution. Since over 70\% of the MIDI files in Lakh belongs to  \textit{pop}, the proportion of \textit{pop} is decreased to around 40\% in Armor, which still remains the greatest, to balance the proportion among genres. Moreover, we increase the proportion of \textit{classical} to the second greatest to ensure the presentation of single track music, which makes up a significant amount of \textit{classical}s from Lakh. For the other 19 genres, we adjust their proportions to balance the ratio between single track and multi-track and between simple and complex music.

 During the labeling process, we ask both professionals and hobbyists to score the comparison and distinguishing tasks so that the general aesthetic criteria of music and expert opinions can both be reflected in our evaluation process. Professionals are people with advanced music theory knowledge while hobbyists do not necessarily possess enough music theory knowledge but have a certain musical connoisseurship. We design two tasks, distinguishing and comparison, for \textit{Armor} since we believe that tasks represents two most common generative tasks. 

\subsection{Models}
\label{sec:models}

\begin{table}[t]
\centering

\begin{tabular}{lrrr}
\toprule
Models & Complexity scores & Multi-track\\
\midrule
MuseGAN & 61.99 & Yes\\
REMI & 84.41 & No\\
AIVA & \textbf{94.45} & Yes\\
Music Transformer & 68.91 & No\\
MuseNet & 74.13 & Yes\\
\midrule
Standard deviation & 11.47 & -\\
\bottomrule
\end{tabular}
\caption{Pilot study on complexity levels and generation tracks of different generative models.}
\label{tab:pilot}
\vspace{-5pt}
\end{table}

We select five most representative and advanced models from recent years to generate music for \textit{Armor}: REMI \cite{REMI}, MuseNet \cite{christine2019musenet}, AIVA, Music Transformer \cite{huang2018musictransformer} and MuseGAN \cite{dong2017musegan}. Except AIVA, the access to whose specific model structures is yet unavailable, the models above are representatives of the most common approaches in music generation: Transformer and GAN. Also, we choose these models to balance single track and multi-track music in \textit{Armor}. Since we want to incorporate music with a wide range of structure complexities, we could only choose models so that each model represents a distinct musical complexity level. To assure that, we calculate the rhythmic complexity for the excerpts generated from generative models using note onsets cross-correlation \cite{yodfat2020complexity} and perform a pilot study to test the musical complexity level for the generative models. The results are shown in Table \hyperref[tab:pilot]{2}. We could tell from the table that the complexity levels of music are evenly distributed in the interval of 60 to 100, which meet our requirements to the distribution of their complexity levels. 

\begin{figure}[h]
  \centering
  \includegraphics[width=\linewidth]{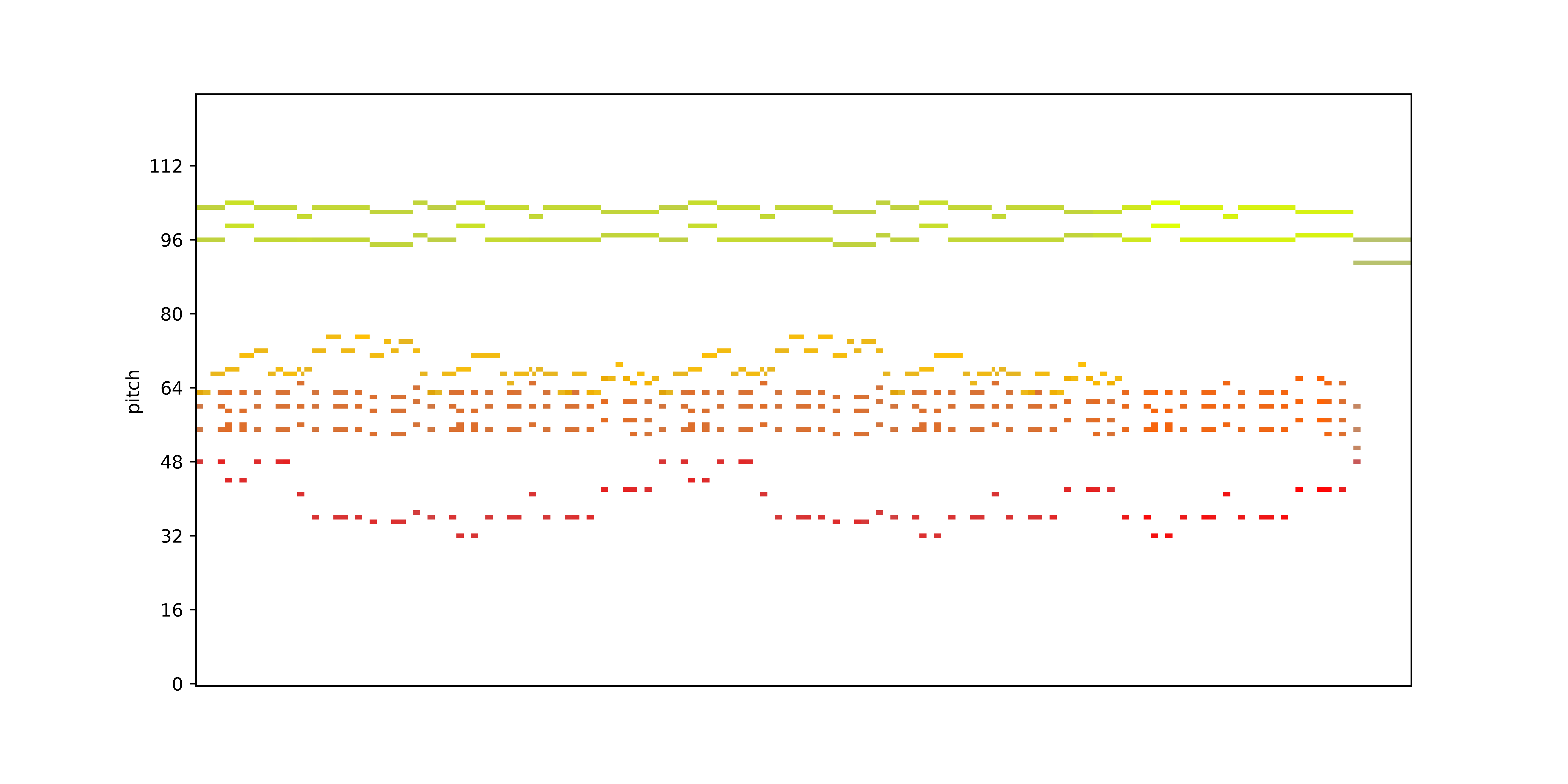}
  \label{fig:aiva}
  \caption{Music generated from AIVA}
\end{figure}

\subsection{Music generation from scratch}
We use all the five models to generate excepts of music from scratch (Example shows in Figure \hyperref[fig:aiva]{2}.). Then we ask our experts, who received at least five years of musical training, to manually classify the generated excerpts into \textit{21} genres. Afterwards, we incorporate same number of pieces of music chosen from Lakh \cite{raffel2016lakh} into the corpus according to the genre and instruments distributions of generated music. Finally, we give each person 20 pieces of music and ask 30 professionals and 30 hobbyists to perform Turing-like test\footnote{This experiment is not really a Turing test, since interaction does not come into play. Hence, we refer to the experiment as a Turing-like test.} on these music. They will provide a binary label indicating whether they believe the music is composed by a human.

\begin{figure}[h]
  \centering
  \includegraphics[width=\linewidth]{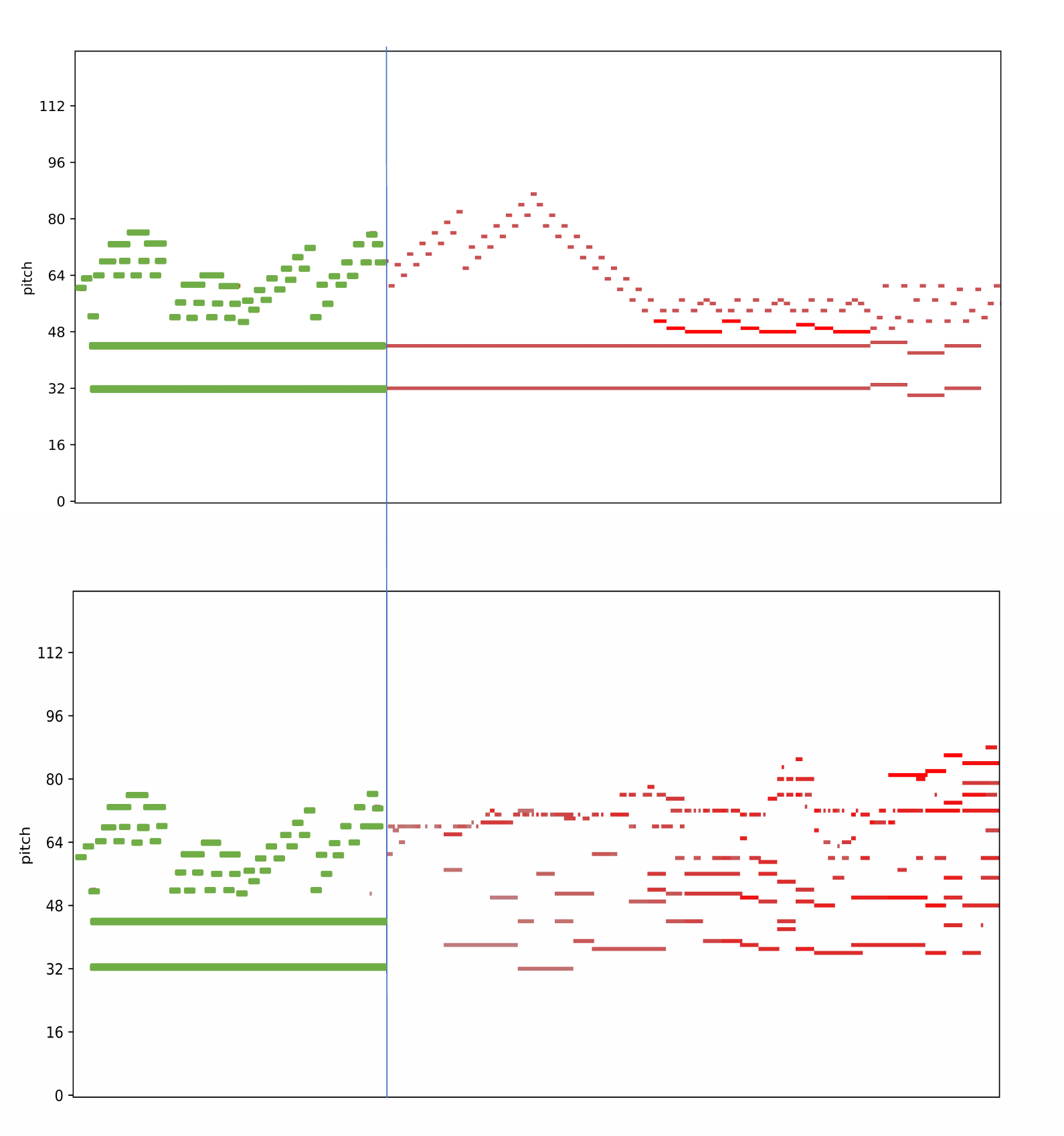}
  \caption{Generated continuation pieced (the upper figure) from REMI compared to its original piece (the lower figure). The notes in green, the prompt, are identical but the red notes diverge.}
  \Description{Continuation compare}
  \label{fig:continuation}
\end{figure}

\subsection{Continuation}

For data collection of the comparison task, we use MuseNet, REMI and Music Transformer to generate continuations of prompts. We excerpted the 30\% of the original piece from the beginning into a prompt because the first 30\% of a composition usually includes its introduction and the beginning stage of establishing musical ideas, which provides the model with adequate information of the music piece without definitive musical patterns. Then we feed the prompt into the model to generate a continuation piece and concatenate the continuation part with the prompt to make a complete musical piece. An example of REMI's continuation is demonstrated by Figure \hyperref[fig:continuation]{3}. Since the model gains some information such as chord progressions and melodies from the prompt, the continuation piece will inherit characteristics of the original piece to some extent and thus share some similarities with the original piece. After we generate enough continuation pairs, we ask both hobbyists and professionals to label the corpus a score between 0 to 1, where 0 suggests the generated piece share no similarities with the original piece and 1 suggests that the generated piece is identical to the original one. Each continuation pair gets 3 scores, two labeled by hobbyists and one labeled by a professional. We abandon AIVA and MuseGAN for this task because: 1) AIVA is not capable of generating continuations 2) MuseGAN has a poor performance on interpolation tasks. The average complexity score for the continuation piece is below 50, which nearly creates no difficulties for human and OE metrics to recognize.

\section{Dataset analysis}




In this section, we present the statistics of \textit{Armor} and show how the diversity of our dataset in genres, instruments and number of tracks would affect the performance of human evaluators. 

\subsection{Armor}

\begin{figure}
  \includegraphics[width=9cm]{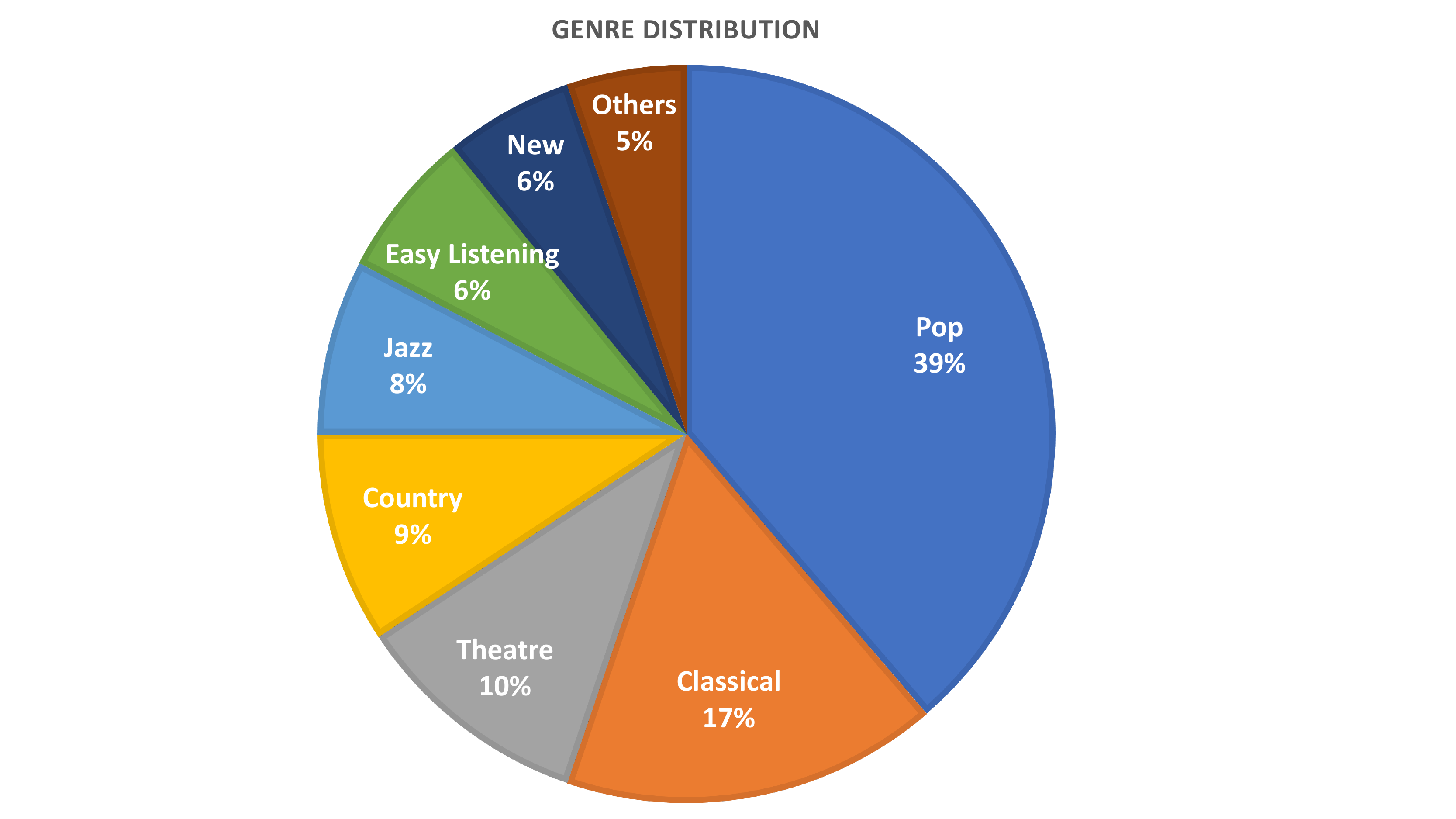}
  \caption{Distribution of genres}
  \label{fig:genres}
\end{figure}

\begin{figure}
  \includegraphics[width=8cm]{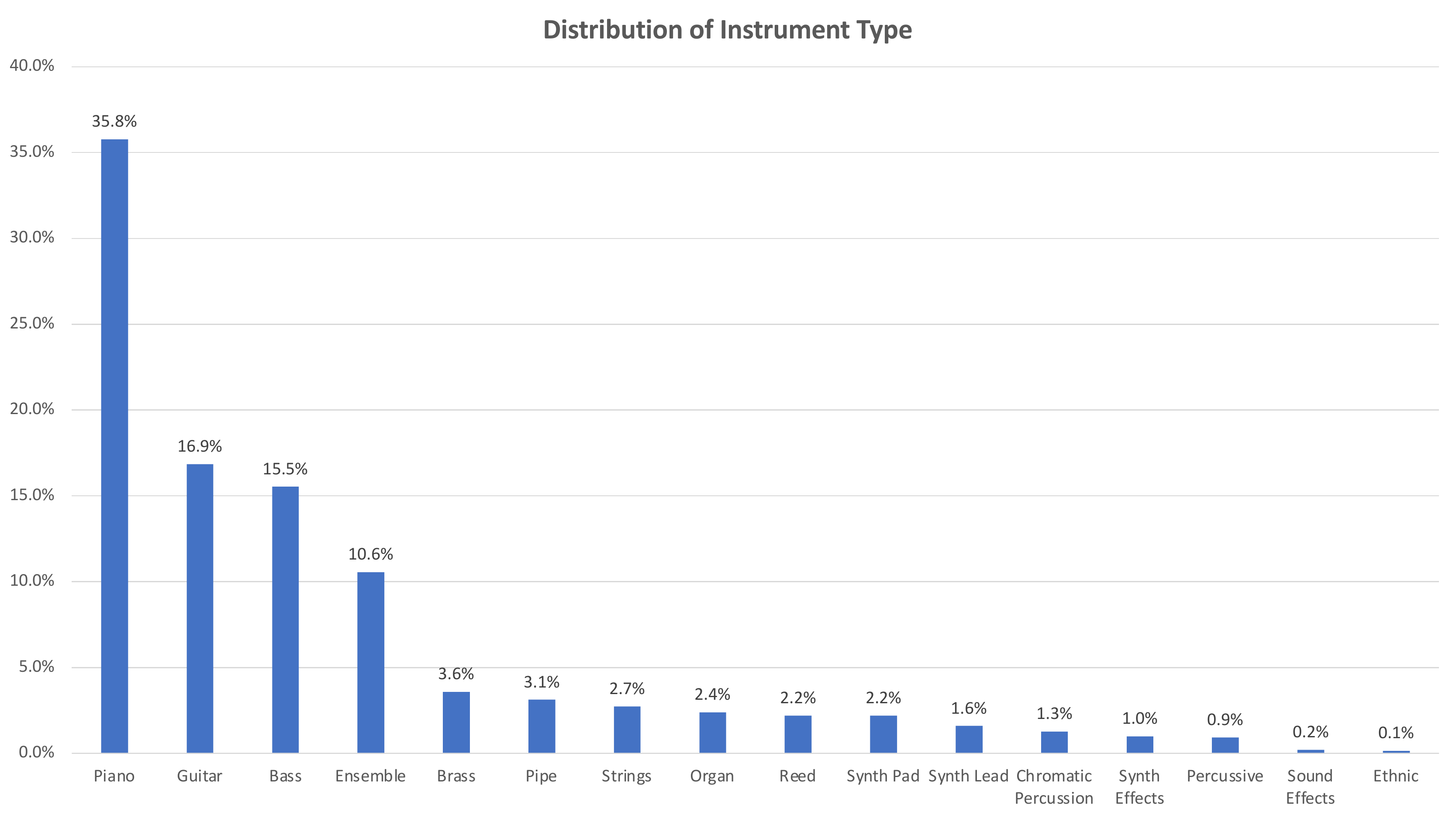}
  \caption{Distribution of instruments}
  \label{fig:instruments}
\end{figure}

We collect \textit{574} pieces of music for the distinguishing task and \textit{248} pairs (\textit{496} pieces of music) for the comparison task. Among these music pieces, there are \textit{21} genres distributed according to the pie chart in Figure \hyperref[fig:genres]{4}. Pop and classic are the top two genres that combined consist of over half of the dataset, and the 14 genres with the least proportions only consist 5\% of the whole datasets. Nevertheless, in the original Lakh \cite{raffel2016lakh} dataset, pop counts over 70\% of the dataset. In terms of instruments, we classify the original 128 types of instruments in midi files into 16 types and plot the distribution as in Figure \hyperref[fig:instruments]{5}. We could tell from the figure that piano, guitar, bass and ensemble count for 78.8\% of the overall instruments and these four families of instruments are commonly used when composing multi-track music. And multi-track music counts for 53.23\% while single-track music counts for 46.77\% of the dataset.

\begin{table}[t]
\centering
\begin{tabular}{lccc}
\toprule
Model/Genre/Track & Hobbyists & Professionals & Total

\\
\midrule
MuseNet & 59.65\% & 86.92\% & 73.29\% \\
AIVA & 52.94\% & 82.81\% & \textbf{67.88\%}\\
Music Transformer & 56.82\% & 86.36\% & 71.59\%\\
REMI & 58.02\% & 85.19\% & 71.61\%\\
MuseGAN & 90.24\% & 97.56\% & \textbf{93.90\%}\\
All & 62.26\% & 87.02\% & 74.64\%\\
\midrule
Pop & 64.71\% & 86.67\% & 75.69\%\\
Classical & 51.72\% & 80.00\% & 65.86\%\\
Theatre & 55.76\% & 82.93\% & 69.85\%\\
Country & 61.10\% & 85.00\% & 73.06\%\\
Others & 64.06\% & 88.11\% & 76.09\%\\
\midrule
Single-track & 56.00\% & 80.86\% & 68.43\%\\
Multi-track & 68.18\% & 90.40\% & 79.29\%\\
\bottomrule
\end{tabular}
\caption{Summary of hobbyists and professionals accuracy on different models, genres and number of tracks for Turing-like test. AIVA outperforms MuseGAN considerably.}
\label{tab:turinganalysis}
\end{table}

\subsection{Distinguishing}
We performed two sets of Turing-like tests on two types of subjects: music hobbyists and professionals with advanced music theory knowledge. Table \hyperref[tab:turinganalysis]{3} shows the accuracy of hobbyists and professionals concerning genres, models and track properties. Unsurprisingly, the professionals outperformed the hobbyists by accurately labeling 87.02\% of the 574 excerpts on whether the excerpt was composed by human or AI, while hobbyists scored an accuracy of 62.26\% on a set of 514 excerpts. The generated music could easily deceive music hobbyists. However, professionals are significantly more sensitive to over-repetition and deliberate complexity, thus less likely to be tricked by model-generated music. Therefore, the models still have considerable space for improvements before they could generate satisfying music.

We also find that all subjects have better accuracy when labeling multi-track compositions than single-track ones, which suggests that models produce more human-like music when using less instruments. We believe that compositions with more instruments are more difficult to realize. Therefore, while a model has limitations in composing human-like music, just like a human beginner composer lacks composing skills on writing a symphony, composing single-track music using instruments, such as piano, a wind or brass solo, or drums, is more likely to be human-like. Moreover, we analyze subject’s performance on different genres. Pop, classical, theatre and country are largest categories, and ‘others’ include genres with smaller proportion, such as Jazz, R\&B and ballet. For both hobbyists and professionals, they have a better accuracy on pop and other genres than their average performance of 62.26\% and 87.02\%, respectively. Thus, we consider that subjects do not perform as well on genres including classical, theatre, country, because of a higher percentage of single-track excerpts in those genres, especially classical.

\subsection{Comparison}

Similar to our Turing-like test, we recruit amateur music hobbyists and professionals in the music field to subjectively judge the performance of the models’ capability of composing music based on a short excerpt as an input (referred to as continuation) by grading on the degree of similarity between original and continuation excerpts. Based on the overall statistics of the evaluation process displayed in Table \hyperref[tab:continuationanalysis]{4}, the professionals and hobbyists provide 248 pairs of compositions with an average score of 45.54\%. Unlike the considerable difference in performance from the Turing-like test, both types of scorers do not seem to yield a significant difference when evaluating the similarity between two excerpts.

On average, scorers provide better feedback on multi-track continuations, which suggests that the judges tend to feel that multi-track continuations are more similar to original compositions. Instead of claiming that models write better continuations for multi-track music, we attribute the lower scores of single-track compositions to the fact that human are more eligible to detect gaps between the continuation and original excerpt when listening to only one instrument rather than multiple. In other words, if only one instrument is present, the variations of the continuation are enhanced since no other instruments would compensate for those deviations.

\begin{table}[t]
\centering
\begin{tabular}{lccc}
\toprule
Model/Genre & Hobbyists & Professionals & Total

\\
\midrule
MuseNet & 0.501 & 0.488 & 0.497 \\
REMI & 0.375 & 0.369 & 0.373\\
Music Transformer & 0.464 & 0.455 & 0.461\\

\midrule
Pop & 0.442 & 0.426 & 0.437\\
Classical & 0.494 & 0.481 & 0.489\\
Theatre & 0.462 & 0.473 & 0.465\\
Country & 0.683 & 0.667 & 0.678\\
Others & 0.452 & 0.443 & 0.449\\
\midrule
Single-track & 0.411 & 0.404 & 0.409\\
Multi-track & 0.499 & 0.485 & 0.494\\
\midrule
\midrule
Average score & 0.459 & 0.449 & 0.455\\
St.dev & 0.164 & 0.180 & 0.164\\
\bottomrule
\end{tabular}
\caption{Summary of hobbyists and professionals' similarity scores on different models, genres and number of tracks for continuation test.}
\label{tab:continuationanalysis}

\end{table}

\section{OE methods analysis}

To explore how OE algorithms correspond with SE when comparing original and continuation compositions, we compute the correlation between the similarity scores provided by humans and algorithms. We observe significant distance between objective metrics and human judgment, which proves the value of our work.

In addition to calculating the correlation between the scores given by objective metrics and by all types of evaluators, we also analyze the correlation between OE on different models and genres. With our observations, we propose insights for establishing more reliable, OE metrics.

\subsection{OE Methods}
To analyze the correlation between the similarity scores provided by human and algorithms, we select four OE metrics. Mir\_eval \cite{raffel_mir_eval} and Mgeval \cite{Yang_Lerch_2020mgeval} are both feature-based metrics that calculate similarity based on extracted deep-level musical features. As a method that has been widely adopted, mir\_eval provides similarity measurement on each extracted feature, with which we obtain a processed similarity score followed by applying a weighted average based on the importance of each feature. With features extracted by Mgeval, a more recent and growing evaluation system, we obtain high-dimensional feature vectors\footnote{Out of the nine features that Mgeval can extract, we select the five features represented by a single number. The four removed features are multidimensional, and adding them nearly halve the correlation. } and calculated the similarity score with the cosine similarity between the vectors of features. Correspondingly, non-feature-based algorithms analyze more accessible musical information such as pitch patterns and rhythm. We implement BLEU score, the traditional method for evaluation of information in discrete sequences, along with calculating the area between the melody curves of the original and continuation compositions. The melody curves are shifted horizontally and vertically to minimize the enclosed area between them, which represents a concise and practical geometric measurement of musical similarity from the perspective of regarding music sequences as curves. With the discussed procedures, we establish a pipeline for evaluating the effectiveness of objective metrics on measuring musical similarity.

In order to examine the performance of OEs on distinguishing task, we cluster the same set of MIDI files used for Turing-like test based on the features extracted by Mgeval. We choose Mgeval or Mir\_eval because the features extracted by mir\_eval are suitable for calculation of similarity and need further processing for clustering, whereas Mgeval yields features directly usable for clustering. We implement k-means clustering algorithm based on Euclidean distance and produce binary clustering groups, representing the OE’s judgment on whether the piece is composed by models or human.

\begin{table}[t]
\centering

\begin{tabular}{lcccc}
\toprule
 & Mir\_eval& Mg\_eval & Area & BLEU\\
 \hline\hline

Hobbyists & 0.084 & 0.100 & 0.246 & 0.222\\
Professionals & 0.121 & 0.094 & 0.238 & 0.188\\
Total & 0.100 & 0.101 & 0.251 & 0.217\\
 \hline\hline
MuseNet & 0.050 & 0.087 & 0.206 & 0.171\\
REMI & 0.091 & 0.032 & 0.212 & 0.188\\
Music Transformer & 0.147 & 0.291 & 0.341 & 0.389\\
 \hline\hline
Pop & 0.137 & 0.115 & 0.273 & 0.315\\
Classical & 0.123 & 0.047 & 0.357 & 0.191\\
Theatre & 0.086 & 0.103 & 0.149 & 0.197\\
Country & 0.118 & 0.067 & 0.292 & 0.276\\
Others & 0.160 & 0.143 & 0.276 & 0.193\\
\bottomrule
\end{tabular}

\caption{Correlation between similarity scores given by human evaluators versus evaluation metrics on different evaluators, models and genres.}
\label{tab:metricsperf}
\end{table}

\subsection{Analysis}
According to results displayed in Table \hyperref[tab:metricsperf]{5}, we observe that all four objective metrics have correlation coefficient less than 0.3 over all types of evaluators, which implies relatively weak correlation with human judgment. In addition, after analyzing the correlation between subjective and objective similarity scores across various models, genres, track properties, our findings consistently suggest that the enclosed area between melody curves, which seems to be the least complicated algorithm among all featured metrics, however, has the strongest correlation with human intuition. 

Comprehensively, the correlation coefficients of human evaluation and algorithms on Music Transformer, which generates only single-track continuations, are the highest. Classical music, which has the highest proportion of single-track compositions, also witnesses the peak performance of the objective metrics across all genres, which implies the outperformance of objective metrics in single-track over multi-track continuation compositions. We believe that features and melodies can be more accurately extracted from single-track excerpts. Thus, such observation suggests that implementing OE metrics could be more reasonable and reliable for comparison of single-track compositions. 

Similar to the inadequate correlation between OEs and human performance on comparing task, the result of clustering also illustrates significant disparity between distinguishing music based on extracted features versus human perception: the clustering method yields an accuracy of only 50.4\%, nearly as low as the theoretical accuracy of 50\% by random guess. Such result demonstrates the inability for OEs to extract ideal features for acceptable clustering results, thus perform well on distinguishing task. 

It is noticeable that the practicality and validity of non-feature-based metrics are thoroughly undermined by feature-based ones throughout all dimensions of analysis on models, genres and track properties, which demonstrates the disadvantages of feature extraction when matching human measurement of musical similarity. We deem the features extracted with OEs insufficient to comprehensively represent the content of symbolic music in MIDI format, which leads to the OE’s the poor correlation on comparison task and the unsatisfactory accuracy on the distinguishing task. 
\subsection{Prospects}
We think that feature extractions inspect deep-level musical information but might ignore information that is exclusively obtainable from a more macroscopic perspective of human evaluation. Since human evaluators perceive musical information as they hear new notes, instead of summarizing the entire excerpt afterwards, evaluation metrics that analyze music as a whole, such as melody analysis implemented by enclosed area and BLEU, might resemble the process of human evaluation to a greater extent. Whereas unlike enclosed area, BLEU approaches the melody by inspecting smaller sequences within the excerpt. BLEU's slightly lower correlation further implies the potential of analyzing melody integrally. Therefore, we propose that in order to devise more effective objective metrics, one could consider methods that are non-feature-based, more intuitive, and are more similar to human process of perceiving musical information. In addition, we encourage other musical elements, such as rhythm, to be taken into consideration.

\section{Conclusion and discussion}
In this paper, we propose a complex and cross-domain benchmark dataset, \textit{Armor}, as a meta-evaluation system to test the effectiveness of music OE metrics. We analyze how genres, models, and single-track/multi-track would affect human evaluators' judgments and the performance of OE metrics. We found that current OE algorithms do not work very well on our dataset. The scores we gained from four evaluation metrics all show weak positive correlations with the human-labeled scores. We discover that evaluation metrics work better on single-track music than on multi-track music, probably because of more accurate extraction of different musical patterns on single-track music. We also found that non-feature-based OE metrics may work better than feature-based methods since non-feature-based methods are closer to human's habit of music appreciation. 

In future works, we hope that 1) we could implement more generative models to incorporate more musical patterns of artificial music 2) we want to test more OE metrics to further prove that our benchmark dataset poses a great challenge for these metrics and gain more insight of how to design a more useful OE metric. 3) Due to limited monetary and human resources, our dataset is relatively small; therefore, we want to enlarge our dataset so that the algorithms tested on our dataset could encounter more test cases. 
Nevertheless, we hope that our work could pave the way for better OE metrics and provide a guideline for future corpus construction.


\begin{acks}
We want to thank CFY, CJL, FWY, YJY, WKD, XRB, CHL, VZ, PYQ, ZYS, DW, ZYM, YXJ, YWF, NZY, FJY, LGX, XYC, MPY, HW, CX, DZY, CR, ZMZ, ZMQ, ZYH, WPY, YYC, BG, XHY, EQH, JTF, QP for their participation in our evaluation process. 
\end{acks}

\clearpage

\bibliographystyle{ACM-Reference-Format}
\balance
\bibliography{citation.bib}


\end{document}